\documentclass[conference]{IEEEtran}
\ifCLASSINFOpdf
\else
\fi
\usepackage{algorithmic}

\usepackage{algorithm}
\usepackage{graphicx}
\usepackage{cite}
\begin{document}
%
\title{An iterative merging placement algorithm for the fixed-outline floorplanning}

\author{\IEEEauthorblockN{Kun He}
\IEEEauthorblockA{School of Computer Science\\
and Technology\\Huazhong University of Science\\
and Technology\\Wuhan, China 430074\\
Email: brooklet60@gmail.com}
\and
\IEEEauthorblockN{Pengli Ji}
\IEEEauthorblockA{School of Computer Science\\
and Technology\\Huazhong University of Science\\
and Technology\\Wuhan, China 430074\\
Email: jipengli8@gmail.com}
\and
\IEEEauthorblockN{Chumin Li}
\IEEEauthorblockA{School of Computer Science\\
and Technology\\
University of Picardie Jules Verne\\
Amiens, France 80039\\
Email: chu-min.li@u-picardie.fr}}


%


\maketitle

\begin{abstract}

Given a set of rectangular modules with fixed area and variable dimensions, and a fixed rectangular circuit. The placement of Fixed-Outline Floorplanning with Soft Modules (FOFSM) aims to determine the dimensions and position of each module on the circuit. We present a two-stage Iterative Merging Placement (IMP) algorithm for the FOFSM with zero deadspace constraint. The first stage iteratively merges two modules with the least area into a composite module to achieve a final composite module, and builds up a slicing tree in a bottom-up hierarchy. The second stage recursively determines the relative relationship (left-right or top-bottom) of the sibling modules in the slicing tree in a top-down hierarchy, and the dimensions and position of each leaf module are determined automatically. Compared with zero-dead-space (ZDS) algorithm, the only algorithm guarantees a feasible layout under some condition, we prove that the proposed IMP could construct a feasible layout under a more relaxed condition. Besides, IMP is more scalable in handling FOFSM considering the wirelength or without the zero deadspace constraint.

\end{abstract}


%
\IEEEpeerreviewmaketitle

\section{Introduction}

Floorplanning is a critical phase in the physical design of VLSI circuit. Nowadays, with the Integrated Circuit (IC) technology advance, design complexity is growing in a dramatic speed, and floorplanning remains to be a difficult problem. Given a set of rectangular modules and a netlist specifying the interconnection among the modules, the floorplanning asks to orthogonally place all the modules onto a rectangular circuit without overlapping, such that the area of the enveloping rectangle that exactly encloses all the modules and the total wirelength of the netlist are minimized. Under the wide utilization of the hierarchial methodology in the VLSI design \cite{classicalharmful}, the outline of the circuit is usually specified beforehand, and the fixed-outline constraint has become a common feature for the floorplanning.

The modules in floorplanning can be classified into hard module and soft module. Hard module is in the fixed dimensions, while the soft module is in the fixed area but variable dimensions. Many algorithms have been developed in the past decades for the floorplanning with the two kinds of modules.

The most classic method to handle floorplanning with hard modules is to combine the representation of the geometric relationships \cite{slicingtree,sp,o-tree,b-tree} among the modules with simulated annealing (SA) \cite{SA1,SA2}. In order to handle large scale modules, hierarchical and multilevel methodologies are adapted then. By scalably incorporating legalization into hierarchical flow, cong et al. \cite{PATOMA} proposed a partitioning to optimize module arrangement (PATOMA) algorithm, which utilized two placement algorithms zero-dead-space (ZDS) and row-oriented block (ROB) to guarantee the legalization of each partitioning. Based on the principle of Deferred Decision Making (DDM), Yan et al. \cite{DeFer} presented an efficient algorithm DeFer. Chan et al. \cite{newbest} introduced a flexible flow to handle floorplanning with mixed modules, and there were two stages in their method: the global distribution stage aims to obtain shorter wirelength while distributing modules over the fixed-outline and the legalization stage aims to obtain feasible solution.

As the aspect ratio (the height divided by the width) of soft module varies continuously, the analytical method \cite{Luo,ZDS,he,SKB,SDS} is the most effective algorithm to handle floorplanning with soft modules. Luo et al. \cite{Luo} introduced a nonlinear optimization methodology. First, it adapted a convex optimization to globally minimize the wirelength globally. Then, a further optimization and legalization was conducted by sizing the aspect ratio of the modules. Based on a recursive top-down area bipartitioning, Cong et al. \cite{ZDS} suggested a zero-dead-space (ZDS) placement algorithm for floorplanning with zero deadspace constraint. He et al. \cite{he} and Lin et al. \cite{SKB} proposed two representations, Ordered Quadtree and SKB-Tree, to encode the layout respectively. Given a topological structure, the above two methods could determine a corresponding layout by analytical approaches. Yan et al. \cite{SDS} developed an optimal slack-driven block shaping (SDS) algorithm to shape the soft modules such that the resulting layout is inside the fixed outline.

In this paper, we present a two-stage Iterative Merging Placement (IMP) algorithm for the Fixed-Outline Floorplanning with Soft Modules (FOFSM) and zero deadspace constraint. At the first stage, all the modules are merged into one composite module by iteratively merging two modules with the least area. For each composite module generated in the first stage, the merging direction and relative position of the two sub-modules have not specified. Then at the second stage, the final composite module is placed on the circuit, and the dimensions and positions of each pair of sub-modules are recursively determined basing on the aspect ratio of the composite module. We then present a mathematic analysis which shows IMP can place all the modules feasibly under some condition. Compared with ZDS, the only algorithm guarantees a feasible layout, the condition of IMP is more relaxed and IMP is more scalable in handling FOFSM considering the wirelength or having zero deadspace constraint.

The remainder of this paper is organized as follows. Section \uppercase\expandafter{\romannumeral2} presents the problem statement. Section \uppercase\expandafter{\romannumeral 3} gives the definition of the composite module, and Section \uppercase\expandafter{\romannumeral 4} describes the Iterative Merging Placement (IMP) algorithm and analyzes conditions for feasible placement. The comparison between IMP and ZDS is discussed in Section \uppercase\expandafter{\romannumeral 5}. And the paper is concluded in Section \uppercase\expandafter{\romannumeral 6}.

\section{Problem Formulation}

Given a set of $n$ soft modules with each module $m_i$ has a fixed area $s_i$ and an aspect ratio interval $[1/\lambda_i,\lambda_i]$ ($\lambda_i \ge 1$), and a fixed circuit with width $W$ and height $H$. The placement of Fixed-Outline Floorplanning with Soft Modules (FOFSM) aims to determine the dimensions and position of each module on the circuit. Here $\lambda_i$ is the bounding factor of $m_i$. Let $(x_{i1},y_{i1})$ and $(x_{i2},y_{i2})$ denote the coordinates of the bottom-left and upper-right vertices of $m_i$, let width $w_i$ and height $h_i$ represent the two dimensions of $m_i$. Then, the layout for the placement of FOFSM should satisfy the following constraints.

(1) $x_{i2}-x_{i1} = w_i$, $y_{i2}-y_{i1} = h_i$

(2) $1/\lambda_i \le h_i/w_i \le \lambda_i $

(3) $max(x_{i1}-x_{j2}, y_{i1}-y_{j2}, x_{j1}-x_{i2}, y_{j1}-y_{i2}) \ge 0$

(4) $0 \le x_{ik} \le W$, $0 \le y_{ik} \le H$, $k \in \{1,2\}$

Here, $i$, $j$ applies to $1,2,\ldots,n$ ($i \ne j$). Constraint (1) implies that each module should be placed orthogonally on the circuit; constraint (2) gives the aspect ratio interval of each module; constraint (3) indicates that all the modules should be placed with no overlapping and constraint (4) means all the modules should be completely placed on the circuit.

\section{composite module}

Similar to the composite pattern in software engineer, we define a conception of composite module in this section, and analyze its characters.
\newtheorem{definition}{Definition}
\begin{definition}[composite module]
A single soft module is a special composite module. Two composite modules could be merged horizontally or vertically to form a bigger composite module with their two pasted sides having the equal length, and the two modules are called the sub-modules of the bigger composite module .
\end{definition}

\begin{figure}[!t]
\centering
\includegraphics[width=1in]{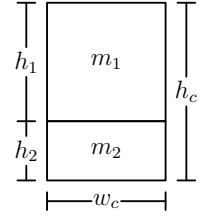}
\caption{composite module.}
\label{fig_sim}
\end{figure}

Fig. 1 illustrates a composite module merged by $m_1$ and $m_2$.

\begin{definition}[feasible composite module]
A composite module is a feasible composite module only if its aspect ratio interval covers $[1/\lambda,1]$ or $[1,\lambda]$ ($\lambda \ge 1$) and its sub-modules are feasible composite modules. And the two sub-modules can be merged feasibly.
\end{definition}

\newtheorem{lemma}{Lemma}
\begin{lemma}
Two sub-modules, which can be merged into one feasible composite module having aspect ratio interval covers $[1/\lambda,1]$ or $[1,\lambda]$, can construct any composite modules having aspect ratio is in $[1/\lambda,\lambda]$ feasibly.
\end{lemma}

\begin{lemma}
suppose there are two module $m_1$ and $m_2$ whose bounding factors are $\lambda_1$ and $\lambda_2$ respectively. Let $\alpha = s_1 / s_2$ ($\alpha \ge 1$), if $1/(\lambda_1-1) \le \alpha \le \lambda_2-1$, then they can merge into a feasible composite module $m_c$ whose bounding factor $\lambda_c \ge min(\lambda_1,\lambda_2)$.

\end{lemma}

\begin{IEEEproof}
Without loss of generality, assume $m_1$ and $m_2$ are merged vertically, as shown in Fig. 1. Let $h_i$ denotes the height of $m_i$, $w_c$ and $h_c$ represent the two dimensions of the composite module $m_c$. Since the bounding factors of $m_1$ and $m_2$ are $\lambda_1$ and $\lambda_2$ respectively

\begin{center}
$$\frac{1}{\lambda_1} \le \frac{h_1}{w} \le \lambda_1 ~ \textrm{,} ~ \frac{1}{\lambda_2} \le \frac{h_2}{w} \le \lambda_2$$
\end{center}

\leftline{Thus}

$$max(\frac{1}{\lambda_1h_1},\frac{1}{\lambda_2h_2}) \le \frac{1}{w} \le min(\frac{\lambda_1}{h_1},\frac{\lambda_2}{h_2})$$

\leftline{Therefore}

$$max(\frac{h}{\lambda_1h_1},\frac{h}{\lambda_2h_2}) \le \frac{h}{w} \le min(\frac{\lambda_1h}{h_1},\frac{\lambda_2h}{h_2})$$

As $s_1=\alpha s_2$, $h_1=\alpha h_2$ and $h=(1+\alpha)h_2$. So, the aspect ratio interval of $m_c$ is

\begin{equation}
max(\frac{1+\alpha}{\alpha\lambda_1},\frac{1+\alpha}{\lambda_2}) \le \frac{h}{w} \le min[(1+\frac{1}{\alpha})\lambda_1,(1+\alpha)\lambda_2]
\end{equation}

By Lemma 1, $m_1$ and $m_2$ can be merged feasibly if $[1,\lambda_c]$ is a subset of the aspect ratio interval of $m_c$. Thus the left part of (1) should be no greater than 1.

$$max(\frac{1+\alpha}{\alpha\lambda_1},\frac{1+\alpha}{\lambda_2}) \le 1$$
so $1+\alpha \le \alpha\lambda_1~ \textrm{and} ~1+\alpha \le \lambda_2$, and we get

\begin{equation}
\frac{1}{\lambda_1-1} \le \alpha \le \lambda_2-1
\end{equation}

\leftline{the bounding factor of $m_c$:}

$$\lambda_c=min[(1+\frac{1}{\alpha})\lambda_1,(1+\alpha)\lambda_2]$$

\leftline{since $\alpha \ge 1$, $(1+1/\alpha)\lambda_1 \ge \lambda_1$ and $(1+\alpha)\lambda_2 \ge \lambda_2$}

\begin{equation}
\lambda_c \ge min(\lambda_1,\lambda_2)
\end{equation}

\end{IEEEproof}

\begin{lemma}
Given the aspect ratio and the position of a composite module, if the merging direction and left-right/top-down order of the two sub-modules are specified, then the aspect ratios and positions of the two sub-modules can be determined uniquely.
\end{lemma}

As shown in Fig. 1, if we know $m_1$ and $m_2$ are merged vertically and $m_1$ is above $m_2$. As the aspect ratio of the composite module is known, the value of $w_c$ can be determined, and the two heights $h_1$ and $h_2$ can be determined then. The aspect ratios and positions of $m_1$ and $m_2$ are determined obviously.

\section{Iterative merging placement algorithm}

The Iterative Merging Placement (IMP) algorithm is a two-stage deterministic algorithm. At the first stage, by iteratively merging two modules with the least area, all the modules will finally be merged into one big composite module having the same area as the circuit. For each composite module generated in this stage, IMP just records its two sub-modules, while the merging direction and left-right/top-down order of the sub-modules are not specified. At second stage, first the final composite module is placed on the circuit such that its aspect ratio and position are specified. Then by Lemma 3, the aspect ratios and positions of each pair of sub-modules are determined based on the aspect ratio of the composite module recursively. The pseudo code of IMP is illustrated in Fig. 2.

\begin{figure}[!t]
\centering

\begin{algorithmic}[]
\footnotesize
\REQUIRE
Module sequence $m_1, m_2, \ldots , m_n$ with respective areas $s_1 \ge s_2 \ldots \ge s_n$;

Circuit region R of area $A=\sum_{k=1}^ns_k$ and dimensions $W \times H$;

\STATE Stage \uppercase \expandafter {\romannumeral 1}
\WHILE {current sequence contains more than one module}
\STATE take out the two rearmost modules $m_i$ and $m_j$ in the sequence and merge them into a composite module $m_c$;
\STATE insert $m_c$ in a position $k$ of the sequence such that $s_{k-1} > s_k \ge s_{k+1}$;
\STATE push $m_c$ into stack $S_T$;
\ENDWHILE

\STATE Stage \uppercase \expandafter {\romannumeral 2}

\STATE place the only composite module in the sequence on the circuit;
\WHILE {$S_T$ is not null}
   \STATE pop one composite module $m_c$ having two sub-modules $m_i$ and $m_j$ from stack $S_T$;
   \IF {$\gamma_c \ge 1$}
      \STATE $m_i$ and $m_j$ merged vertically and the larger one is on the top;
   \ELSE
      \STATE $m_i$ and $m_j$ merged horizontally and the larger one is on the left;
   \ENDIF
   \STATE determining the aspect ratios and positions of $m_i$ and $m_j$ by Lemma 3;
\ENDWHILE

\caption{The iterative merging placement (IMP) algorithm.}
\end{algorithmic}
\end{figure}

\begin{lemma}
The area of one composite module generated at step $t$ of the IMP is no larger than the areas of any composite modules generated after step $t$.
\end{lemma}

At each step of stage \uppercase \expandafter {\romannumeral 1} of the IMP, two modules with the least area are token out to be merged into a composite module which will be inserted into the sequence then. Therefore, the two modules selected are not bigger than that in the following steps.

\newtheorem{theorem}{THEOREM}
\begin{theorem} IMP can place all the modules feasibly under the following conditions,

(1) all the modules have an uniform bounding factor $\lambda$ ($\lambda \ge 3$).

(2) the aspect ratio of the circuit $\gamma_A$ is in the period $[1/\lambda,\lambda]$.

(3) for all $i \in {1,2,\ldots,n-1}$,
   $$max \frac{s_i}{sum_{j=i+1}^n s_j} \le \lambda-1$$

\end{theorem}

\begin{IEEEproof}
Since the bounding factors of all the given modules are $\lambda$, by Lemma 2 all the feasible composite modules generated in the IMP should have bounding factors no smaller than $\lambda$. Thus, to insure IMP can obtain a feasible layout, we just need to prove all the generated composite modules are feasible.

As $\lambda \ge 3$, the left side of the constraint $1/(\lambda_1-1) \le \alpha \le \lambda_2-1$ in Lemma 2 is always true. Thus, to insure $m_1$ and $m_2$ can merge feasibly, we just need to prove $\alpha \le \lambda_2-1$.

Let $c_i$ represents a composite module, $c_i^j$ denotes that module $c_i$ is located in position $j$ in the current sequence. A mathematical induction is used here to prove the composite module generated at each step is feasible.

At step $T=1$. $m_{n-1}$ and $m_n$ are going to be merged. According to condition (1) and (3), $s_{n-1}/s_n  \le \lambda-1= \lambda_n-1 $. Hence they can be merged feasibly.

Then we prove the following assertion: if all the composite modules generated at the first $T-1$ steps are feasible. Then the sub-modules can be merged feasibly at step $T$:

\begin{figure}[!t]
\centering
\includegraphics[width=3.2in]{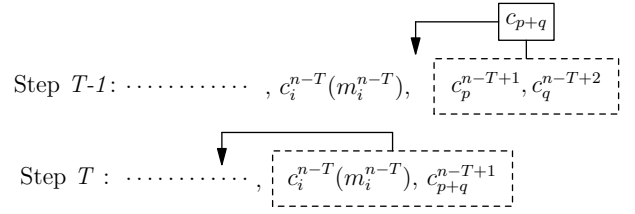}
\caption{The composite module $c_{p+q}$ generated in step $T-1$ insert at the rear of the sequence.}
\label{fig_sim}
\end{figure}

If the composite module $c_{p+q}$ generated at step $T-1$ is inserted at the rear of the sequence, as shown in Fig. 3. By Lemma 4, $c_i$ must be a single module, otherwise $c_{p+q}$ would be inserted in front of $c_i$. Similarly, $c_{p+q}$ is a composite module including $\{m_{i+1},\ldots,m_n\}$, otherwise there should be a step $T_k$ ($1 \le T_k < T-1$) at which a subset of $\{m_{i+1},\ldots,m_n\}$ is merged into a composite module and inserted in front of $c_i$, and $c_{p+q}$ should be inserted in front of $c_i$. Therefore

$$ \frac{s_i}{s_{p+q}} = \frac{s_i}{\sum_{j=i+1}^n s_j}$$

By the assertion $c_{p+q}$ is a feasible composite module, $\lambda_{p+q} \ge \lambda$. Hence

$$ \frac{s_i}{s_{p+q}} \le \lambda-1 \le \lambda_{p+q}-1$$

\begin{figure}[!t]
\centering
\includegraphics[width=2.7in]{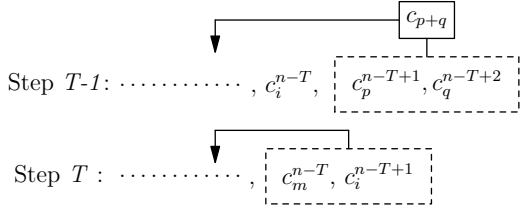}
\caption{The composite module $c_{p+q}$ generated in step $T-1$ does not insert at the rear of the sequence.}
\label{fig_sim}
\end{figure}

Otherwise, the composite module $c_{p+q}$ generated at step $T-1$ is not inserted at the rear of the sequence, as shown in Fig. 4. As the sequence is organized in a non-increasing order, $s_i \ge s_p \ge s_q$, and, $s_{p+q}=s_p+s_q \le 2s_i$. Because $c_{p+q}$ is inserted in front of $c_i$, $s_m \le s_{p+q}$. Therefore

$$ \frac{s_m}{s_i} \le \frac{s_{p+q}}{s_i} \le 2 \le \lambda_i-1 $$

\leftline{Above all, the sub-modules can be merged feasibly at step $T$.}

\end{IEEEproof}

\section{comparison with ZDS}

Based on a recursive top-down area bipartitioning, zero-dead-space (ZDS) \cite{ZDS} is an algorithm attempts to bound the aspect ratios of all the modules uniformly. At each step, the modules in a region are separated into two groups such that the total areas of the groups are as nearly equal as possible. The region is then cut parallel to its shorter side with each group fits exactly into one of the regions. Modules are placed once they fill a sufficient fraction of their subregions. To the best of our knowledge, ZDS is the only method that can deal with the placement of FOFSM under some condition. The condition is that all the given modules have an uniform upper bound $\lambda$ ($\lambda \ge 3$), the aspect ratio of the circuit $\gamma_A$ is in period $[1/\lambda,\lambda]$, and for all $i \in \{1,2, \ldots, n-1 \}$, $max(s_i/s_{i+1}) \le \lambda-1$.

IMP has a more relaxed condition to guarantee a feasible layout. As

$$ max(\frac{s_i}{\sum_{j=i+1}^ns_j}) \le max(\frac{s_i}{s_{i+1}}) $$

Furthermore, IMP is more scalable in handling a related problem, the placement of FOFSM without zero deadspace constraint. As a recursive top-down area bipartitioning method, there is a deadspace distribution at each step of ZDS. Therefore a backtracking to redistribute the deadspace is always necessary when ZDS fails to obtain a feasible layout. By comparison, IMP merges all the modules without the consideration of deadspace distribution, and the aspect ratio intervals of all the composite modules generated in this process can be figured out, which can be used to guide the distributing deadspace to composite modules.

Also, IMP is more scalable in handling another related problem, the FOFSM and considering the wirelength. At each step, The merging operation of IMP is just to merge two modules, while the separating operation of ZDS is to separate a set of modules into two groups and partition the corresponding region. Therefore, the merging operation is simpler, which makes it easy to consider the wirelength optimization.

\section{Conclusion}

In this paper, we deal with a continuous optimization problem, the placement of Fixed-Outline Floorplanning with Soft Modules (FOFSM), which aims to shape and place all the modules on a fixed circuit. Similar to the composite pattern in soft engineer, we proposed a conception of composite module. Based on this conception, an Iterative Merging Placement (IMP) algorithm is suggested to handle the placement of FOFSM. In the IMP, modules with the least area are prior to be merged, and after all the modules are merged, the dimensions and position of each module is determined recursively. We prove that IMP can obtain a feasible layout under a more relaxed condition. Moreover, it is more scalable in handling related problem that has deadspace or considering the wirelength.


\section*{Acknowledgment}

This work was supported by the National Natural Science Foundation of China (Grant no. 61173180 and 61272014).



%
\bibliographystyle{IEEEtran}
\bibliography{myfile}

\begin{thebibliography}{10}
\providecommand{\url}[1]{#1}
\csname url@samestyle\endcsname
\providecommand{\newblock}{\relax}
\providecommand{\bibinfo}[2]{#2}
\providecommand{\BIBentrySTDinterwordspacing}{\spaceskip=0pt\relax}
\providecommand{\BIBentryALTinterwordstretchfactor}{4}
\providecommand{\BIBentryALTinterwordspacing}{\spaceskip=\fontdimen2\font plus
\BIBentryALTinterwordstretchfactor\fontdimen3\font minus
  \fontdimen4\font\relax}
\providecommand{\BIBforeignlanguage}[2]{{%
\expandafter\ifx\csname l@#1\endcsname\relax
\typeout{** WARNING: IEEEtran.bst: No hyphenation pattern has been}%
\typeout{** loaded for the language `#1'. Using the pattern for}%
\typeout{** the default language instead.}%
\else
\language=\csname l@#1\endcsname
\fi
#2}}
\providecommand{\BIBdecl}{\relax}
\BIBdecl

\bibitem{classicalharmful}
A.~B. Kahng, ``Classical floorplanning harmful?'' in \emph{Proceedings of the
  2000 International Symposium on Physical Design}, ser. ISPD '00.\hskip 1em
  plus 0.5em minus 0.4em\relax New York, NY, USA: ACM, 2000, pp. 207--213.

\bibitem{slicingtree}
R.~H. Otten, ``Automatic floorplan design,'' in \emph{Proceedings of the 19th
  Design Automation Conference}, ser. DAC '82.\hskip 1em plus 0.5em minus
  0.4em\relax Piscataway, NJ, USA: IEEE Press, 1982, pp. 261--267.

\bibitem{sp}
H.~Murata, K.~Fujiyoshi, S.~Nakatake, and Y.~Kajitani, ``Vlsi module placement
  based on rectangle-packing by the sequence-pair,'' \emph{Computer-Aided
  Design of Integrated Circuits and Systems, IEEE Transactions on}, vol.~15,
  no.~12, pp. 1518--1524, 1996.

\bibitem{o-tree}
P.-N. Guo, C.-K. Cheng, and T.~Yoshimura, ``An o-tree representation of
  non-slicing floorplan and its applications,'' in \emph{Proceedings of the
  36th Annual ACM/IEEE Design Automation Conference}, ser. DAC '99.\hskip 1em
  plus 0.5em minus 0.4em\relax New York, NY, USA: ACM, 1999, pp. 268--273.

\bibitem{b-tree}
Y.-C. Chang, Y.-W. Chang, G.-M. Wu, and S.-W. Wu, ``B*-trees: A new
  representation for non-slicing floorplans,'' in \emph{Proceedings of the 37th
  Annual Design Automation Conference}, ser. DAC '00.\hskip 1em plus 0.5em
  minus 0.4em\relax New York, NY, USA: ACM, 2000, pp. 458--463.

\bibitem{SA1}
T.-C. Chen and Y.-W. Chang, ``Modern floorplanning based on b*-tree and fast
  simulated annealing,'' \emph{Computer-Aided Design of Integrated Circuits and
  Systems, IEEE Transactions on}, vol.~25, no.~4, pp. 637--650, 2006.

\bibitem{SA2}
S.~Bandyopadhyay, S.~Saha, U.~Maulik, and K.~Deb, ``A simulated annealing-based
  multiobjective optimization algorithm: Amosa,'' \emph{Evolutionary
  Computation, IEEE Transactions on}, vol.~12, no.~3, pp. 269--283, 2008.

\bibitem{PATOMA}
J.~Cong, M.~Romesis, and J.~Shinnerl, ``Fast floorplanning by look-ahead
  enabled recursive bipartitioning,'' \emph{Computer-Aided Design of Integrated
  Circuits and Systems, IEEE Transactions on}, vol.~25, no.~9, pp. 1719--1732,
  2006.

\bibitem{DeFer}
J.~Yan and C.~Chu, ``Defer: Deferred decision making enabled fixed-outline
  floorplanner,'' in \emph{Design Automation Conference, 2008. DAC 2008. 45th
  ACM/IEEE}, 2008, pp. 161--166.

\bibitem{newbest}
K.-C. Chan, C.-J. Hsu, and J.-M. Lin, ``A flexible fixed-outline floorplanning
  methodology for mixed-size modules,'' in \emph{Design Automation Conference
  (ASP-DAC), 2013 18th Asia and South Pacific}, 2013, pp. 435--440.

\bibitem{Luo}
C.~Luo, M.~Anjos, and A.~Vannelli, ``\BIBforeignlanguage{English}{A nonlinear
  optimization methodology for vlsi fixed-outline floorplanning},''
  \emph{\BIBforeignlanguage{English}{Journal of Combinatorial Optimization}},
  vol.~16, no.~4, pp. 378--401, 2008.

\bibitem{ZDS}
J.~Cong, G.~Nataneli, M.~Romesis, and J.~R. Shinnerl, ``An area-optimality
  study of floorplanning,'' in \emph{Proceedings of the 2004 International
  Symposium on Physical Design}, ser. ISPD '04.\hskip 1em plus 0.5em minus
  0.4em\relax New York, NY, USA: ACM, 2004, pp. 78--83.

\bibitem{he}
O.~He, S.~Dong, J.~Bian, S.~Goto, and C.-K. Cheng, ``A novel fixed-outline
  floorplanner with zero deadspace for hierarchical design,'' in
  \emph{Computer-Aided Design, 2008. ICCAD 2008. IEEE/ACM International
  Conference on}, 2008, pp. 16--23.

\bibitem{SKB}
J.-M. Lin and Z.-X. Hung, ``Skb-tree: A fixed-outline driven representation for
  modern floorplanning problems,'' \emph{Very Large Scale Integration (VLSI)
  Systems, IEEE Transactions on}, vol.~20, no.~3, pp. 473--484, 2012.

\bibitem{SDS}
J.~Z. Yan and C.~Chu, ``Optimal slack-driven block shaping algorithm in
  fixed-outline floorplanning,'' in \emph{Proceedings of the 2012 ACM
  International Symposium on International Symposium on Physical Design}, ser.
  ISPD '12.\hskip 1em plus 0.5em minus 0.4em\relax New York, NY, USA: ACM,
  2012, pp. 179--186.

\end{thebibliography}

\end{document}